\documentclass[%
 aip,
 sd,%
 amsmath,amssymb,
 reprint,%
]{revtex4-1}

\usepackage{graphicx}
\usepackage{dcolumn}
\usepackage{bm}

\begin{document}

\preprint{AIP/123-QED}

\title{Effective Stochastic Generator with Site-Dependent Interactions}

\author{Masoumeh Khamehchi}
 \email{m.khamehchi@basu.ac.ir}
\author{Farhad H. Jafarpour}%
 \email{farhad@ipm.ir}
\affiliation{
Physics Department, Bu-ali Sina University, 65174-4161 Hamedan, Iran
}%

\date{\today}

\begin{abstract}
It is known that the stochastic generators of effective processes associated with the unconditioned dynamics of rare events might consist of non-local interactions; however, it can be shown that there are special cases for which these generators can include local interactions. In this paper we investigate this possibility by considering systems of classical particles moving on a one-dimensional lattice with open boundaries. The particles might have hard-core interactions similar to the particles in an exclusion process or there can be arbitrary many particles at single site as in a zero-range process. Assuming that the interactions in the original process are local and site-independent, we will show that under certain constraints on the microscopic reaction rules, the stochastic generator of unconditioned process can be local but site-dependent. As two examples, the asymmetric zero-temperature Glauber model and the A-model with diffusion are presented and studied under the above mentioned constraints.
\end{abstract}

\maketitle

\section{Introduction}
The study of stochastic processes in steady-state both around a typical value of a time-integrated observable or far from it, has become of great interest during recent years. The dynamics of a stochastic process far from the typical value of a defined time-integrated observable can be described by an unconditioned Markov process [\onlinecite{lecomte}-\onlinecite{popkov}]. The stochastic generator of this unconditioned Markov process is, in principle, non-local. It might also contain long-range interactions even if the interactions in the original process are short-range. As an example, this unconditioned dynamics, which is sometimes called the effective dynamics, is studied for a one-dimensional Ising chain with periodic boundary condition in which the model undergoes a phase transition from paramagnetic phase, consisting of short-range interactions, to a ferromagnetic phase resulting from long-range interactions [\onlinecite{jack}]. In [\onlinecite{torkaman}], certain constraints are found for a system of classical particles with hardcore interactions defined on a one-dimensional lattice with open boundaries and nearest-neighbor interactions under which the generator of effective stochastic process has the same dynamical rules (reaction rules) as the original process.

In this paper, in comparison to what has been done in [\onlinecite{torkaman}], we aim to investigate a more general case. By defining a time-integrated physical observable and considering an stochastic Markov process of interacting particles defined on a finite lattice of \emph{L} lattice-sites with open boundaries and assuming that the interactions between the particles are short-range, local and site-independent, we investigate the constraints under which the effective interactions in the effective dynamics of the original system which is conditioned on an atypical value of a time-integrated physical observable, are local and short-range but site-dependent. This means that the reaction rules depend on the lattice-site
where the interaction occurs.

A similar study has been done in [\onlinecite{hirschberg}] for a zero-range process defined on a finite lattice of length \emph{L} with open boundaries. It has been shown that under certain constraints and for an atypical value of the local current between the left reservoir and the first lattice-site of the chain, the interactions in the effective dynamics of the zero-range process are site-dependent. This feature can be quite important since the microscopic interactions influence the macroscopic properties of non-equilibrium steady-states. For example, in [\onlinecite{evans}], it has
been shown that the condensation transition and also the coarsening dynamics in a homogenous zero-range processes (where the transition rates are site-independent) and the heterogenous zero-range processes (where the transition rates are site-dependent) are different.

 As we have already mentioned we will consider systems of classical particles on finite lattices with open boundaries. Two types of particles will be considered: the hardcore particles and noninteracting particles. The hardcore particles have hard-core interactions so that each lattice site can be occupied by at most one of them while noninteracting particles can accumulate on a single lattice site.  We will show that our approach is quite general so that it contains the calculations and results of [\onlinecite{torkaman},\onlinecite{hirschberg}]. It means that their work can be considered as special cases of our calculations. It is worth mentioning that our calculations hold for any arbitrary time-integrated observable which can be local or global. Two types of time-integrated observables are known. The first type observable depends on the transition between consecutive configurations that the system meets during the observation time. The second type observable depends on the configurations that the system visits during the observation time.

Certain constraints will be obtained for both hardcore particle and noninteracting particle systems in the most general form under which the transition rates of the effective stochastic process are site-dependent. We will present two examples by considering the activity of the system (the number of times that the system changes its configuration during the observation time) as time-integrated observable. The first example is the asymmetric zero temperature Glauber model [\onlinecite{masharian}]. The second example is the A-model with diffusion [\onlinecite{garrahan}].  The explicit form of stochastic generator for the effective processes are obtained for both cases. It is shown that the effective generators of both examples under the constraints are short-range and site-dependent.

The paper is organized as follows. In section II, the mathematical tools are introduced and the basic concepts are reviewed briefly. In section III, the constraints are determined under which for each arbitrary time-integrated observable in a hardcore particle system, the transition rates of effective stochastic process are local and site-dependent. We will finally investigate the asymmetric zero temperature Glauber model as an example. In section IV, we determine the constraints under which for each arbitrary time-integrated observable in a noninteracting particle system, the transition rates of effective stochastic process are local and site-dependent. We will finally present the A-model with diffusion as an example.

\section{Mathematical tools}
We start with a stochastic Markov process in continuous time dynamics. This is defined by a finite set of configurations $\{C\}$, and transition rates $\emph{w} (C\rightarrow \acute{C})$ for transitions from configurations $C$ to configurations $\acute{C}$. The probability $p(C,t)$ of finding the system in configuration $\emph{C}$ at time $\emph{t}$ evolves in time according to the following master equation,
\begin{equation}
\label{root101}
\partial_t p(C,t)=- r(C) p(C,t)+\sum_{\acute{C}}w(\acute{C}\rightarrow C)p(\acute{C},t)
\end{equation}
where $r(c)=\sum_{\acute{C}}w(C\rightarrow\acute{C})$ is the rate of escape from configuration $\emph{C}$.

To study the dynamics of stochastic process, we start with a "path microcanonical approach", classifying paths or trajectories (a trajectory is defined by a sequence of configurations and a sequence of times) by the values of a time-integrated observable. Two types of time-integrated observables are known. The first type observable $A_1$ depends on the transitions between consecutive configurations that the system meets during the observation time.
\begin{equation}
\label{root104}
A_1=\sum_{jumps C\rightarrow\acute{C}}a_1(C\rightarrow\acute{C})
\end{equation}
where the sum runs over all transitions within a trajectory. For example, if $a_1(C\rightarrow \acute{C})=1$, then $A_1$ is the activity. $a_1(C\rightarrow\acute{C})$ is the contribution of the stochastic transition $C\rightarrow \acute{C}$ to $A_1$.\\
The second type observable $A_2$ depends on configurations that the system visits during the observation time.
\begin{equation}
\label{root105}
A_2=\int_{0}^{t} d\tau a_2(C(\tau))
\end{equation}
where $C(\tau)$ is the configuration of the system at $\tau$. $a_2(C(\tau))$ depends only the configuration of the system at time $\tau$.
Taking the path microcanonical approach, the probability  $p(C,A_1,t)$ of being at configuration $\emph{C}$ at time $\emph{t}$, having observed a value $A_1$ of the time-integrated observable up to $t$ evolves in time according to the following equation:
\begin{eqnarray}
\label{root204}
\partial_tp(C,A_1,t)&=&\sum_{\acute{C}}w(\acute{C}\rightarrow C) p(\acute{C},A_1-a_1(\acute{C}\rightarrow C),t)\nonumber\\&-&r(C)p(C,A_1,t)
\end{eqnarray}
and the probability $p(C,A_2,t)$ of being at configuration $\emph{C}$ at time $\emph{t}$, having observed a value $A_2$ of the observable up to $t$ evolves in time as:
\begin{eqnarray}
\label{root205}
\partial_tp(C,A_2,t)&=&\sum_{\acute{C}}w(\acute{C}\rightarrow C) p(\acute{C},A_2,t)-(r(C)p(C,A_2,t)\nonumber\\&+&a_2(C)\partial_{A_2}p(C,A_2,t))
\end{eqnarray}

On the other hand, considering a biasing field \emph{s} and moving to the Laplace space of the observable, we can construct a biased dynamics. This is the same as moving from a microcanonical ensemble to a canonical one in equilibrium statistical mechanics. Here, \emph{s} plays the role of inverse temperature in the canonical ensemble.\\
Taking the Laplace transform of \eqref{root204} with respect to $A_1$, the equation of evolution for $p(C,A_1,t)$ is
\begin{eqnarray}
\label{root3333}
\partial_tp(C,s,t)&=&\sum_{\acute{C}} e^{-sa_1(\acute{C}\rightarrow C )}w(\acute{C}\rightarrow C) p(\acute{C},s,t)\nonumber\\&-&r(C)p(C,s,t)
\end{eqnarray}
In an operator notation $\partial_t |P(s,t)\rangle=W(s)|P(s,t)\rangle$ [\onlinecite{scheutz}], where $W(s)$ is the generator of biased dynamics in the space of configurations $\{C\}$, its matrix elements are
\begin{equation}
\label{root106}
(W(s))_{C,\acute{C}}=w(\acute{C}\rightarrow C)e^{-s a_1(\acute{C}\rightarrow C)}-r(C) \delta_{C,\acute{C}}
\end{equation}
As can be seen, only non-diagonal part of $W(s)$ is modified by $s$.\\
Also, we take the Laplace transform of \eqref{root205} with respect to $A_2$, the equation of evolution for $P(C,A_2,t)$ is
\begin{eqnarray}
\label{root3334}
\partial_tp(C,s,t)&=&\sum_{\acute{C}}w(\acute{C}\rightarrow C) p(\acute{C},s,t)-(r(C)+sa_2(C))\nonumber\\& &p(C,s,t)
\end{eqnarray}
 In the operator notation mentioned above, the matrix elements of $W(s)$ are
\begin{equation}
\label{root107}
(W(s))_{C,\acute{C}}=w(\acute{C}\rightarrow C)-(r(C)+s a_2(C)) \delta_{C,\acute{C}}
\end{equation}
This time, it is the diagonal part of the generator of biased dynamics which is modified by $\emph{s}$.

Alternatively, It has been shown that there is a time translation invariant regime during which one can construct an effective (unconditioned) stochastic dynamics whose unbiased trajectories coincides with the those of biased dynamics [\onlinecite{jack}].\\
Considering the eigenvalue equations for $W(s)$
 \begin{equation}
\label{root110}
W(s)|\Lambda (s)\rangle=\Lambda (s) |\Lambda(s)\rangle
\end{equation}
\begin{equation}
\label{root111}
\langle\tilde{\Lambda}(s)|W(s)=\Lambda (s)\langle\tilde{\Lambda} (s)|
\end{equation}
it has been shown that using a generalization of Doob's h-transformation of the generator of biased dynamics $W(s)$, the stochastic generator of unconditioned (effective) process is given by [\onlinecite{jack},\onlinecite{chetrite}]
 \begin{equation}
\label{root112}
W^{eff}(s)=U W(s)U^{-1}-\Lambda^*(s)
\end{equation}
where $U$ is a diagonal matrix with elements $\langle C|U|C\rangle=\langle \tilde{\Lambda}^*(s)|C \rangle$. The asterisk shows the largest eigenvalue and corresponding left and right eigenvectors of the generator of biased dynamics $W(s)$.\\
Also, transition rates of the effective Markov process are [\onlinecite{popkov}]:
\begin{equation}
\label{root113}
(W^{eff}(s))_{\acute{C},C}=(W(s))_{\acute{C},C}\frac{\langle\tilde{\Lambda}^*(s)|\acute{C}\rangle}{\langle\tilde{\Lambda}^*(s)|C\rangle} ~~~for~ C\neq\acute{C}
\end{equation}
in which $(W(s))_{\acute{C},C}$, $(W^{eff}(s))_{\acute{C},C}$ are rates of change $C\rightarrow\acute{C}$ in the biased and in effective stochastic process, respectively.

\section{effective generator with site-dependent interactions in a hardcore particle system}

We consider a system of hardcore particles with nearest neighbor interactions on a finite one-dimensional lattice with open boundaries, from there particles can enter or exit the lattice. The dynamical rules in the original process is
\begin{center}
$00\rightarrow 01$~~~with rate $\omega_{21}$~~;~~$00\rightarrow 10$~~~with rate $\omega_{31}$ \\
$01\rightarrow 00$~~~with rate $\omega_{12}$~~;~~$10\rightarrow 00$~~~with rate $\omega_{13}$ \\
$01\rightarrow 10$~~~with rate $\omega_{32}$~~;~~$10\rightarrow 01$~~~with rate $\omega_{23}$ \\
$01\rightarrow 11$~~~with rate $\omega_{42}$~~;~~$10\rightarrow 11$~~~with rate $\omega_{43}$ \\
$11\rightarrow 01$~~~with rate $\omega_{24}$~~;~~$11\rightarrow 10$~~~with rate $\omega_{34}$\\
$00\rightarrow 11$~~~with rate $\omega_{41}$~~;~~$11\rightarrow 00$~~~with rate $\omega_{14}$ \\       
\end{center}
 We investigate the constraints under which the effective generator of the original system with an atypical value of a time-integrated physical observable contains local (short-range) but site-dependent interactions. Since $U$ in \eqref{root112} is a diagonal matrix with elements $\langle C| U | C\rangle = \langle\tilde{\Lambda}^*(s)| C \rangle$, one choice is where $\langle \tilde{\Lambda}^* (s) |$ is in a product form
\begin{equation}
\label{root12}
\langle\tilde{\Lambda}^*(s)|=(1~~ x_1)\otimes(1~~x_2)\otimes(1~~x_3)\otimes(1~~ x_4)\otimes...\otimes(1~~ x_L)
\end{equation}
Considering an arbitrary time-integrated observable, $W(s)$ cab be written as
 \begin{eqnarray}
\label{root13}
\emph{W(s)}&=&\emph{l(s)}\otimes\emph{I}^{\otimes\emph{L}-1}+\Sigma_{i=1}^{L-1}\emph{I}^{\otimes\emph{i}-1} \otimes \emph{h(s)}\otimes\emph{I}^{\otimes L-\emph{i}-1}\nonumber\\&+&\emph{I}^{\otimes{\emph{L}-1}}\otimes\emph{r(s)}
\end{eqnarray}
in which \emph{I} is a 2 $\times$ 2 identity matrix and $h(s)$, $\emph{l(s)}$, and $r(s)$ are
\[
\emph{h(s)}=\left(
\begin{array}{cccc}
  \acute{\acute{\omega_{11}}} &\acute{ \omega_{12}} & \acute{\omega_{13}} & \acute{\omega_{14}} \\
 \acute{ \omega_{21}} & \acute{\acute{\omega_{22}}} & \acute{\omega_{23}} & \acute{\omega_{24}} \\
  \acute{\omega_{31}} & \acute{\omega_{32}} & \acute{\acute{\omega_{33}}} & \acute{\omega_{34}} \\
  \acute{\omega_{41}} & \acute{\omega_{42}} & \acute{\omega_{43}} & \acute{\acute{\omega_{44}}}
\end{array}\right)
\] \\
the "primed" and "double primed" rates are non-diagonal and diagonal elements of $h(s)$, respectively. They are obtained through \eqref{root106} or \eqref{root107} based on the type of the time-integrated observable that we consider in the calculations. \\
\[
\emph{l(s)} = \left(
\begin{array}{cc}
a &b \\
c & d
\end{array}
\right)
\]
\[
\emph{r(s)} = \left(
\begin{array}{cc}
f & g \\
h &k
\end{array}
\right)
\]
Considering \eqref{root111}, certain constraints are obtained under which $\langle \tilde{\Lambda}^* (s)|$ is in form of \eqref{root12}.
\begin{equation}
\label{root14}
L_1=L_2+C_{12}-A_{12}
\nonumber
\end{equation}
\begin{equation}
\label{root15}
R_1=R_2+B_{L-1,L}-A_{L-1,L}
\nonumber
\end{equation}
\begin{equation}
\label{root16}
B_{i,i+1}+C_{i+1,i+2}=A_{i,i+1}+A_{i+1,i+2}
\nonumber
\end{equation}
\begin{equation}
\label{root17}
A_{i,i+1}+D_{i,i+1}=B_{i,i+1}+C_{i,i+1}
\end{equation}
in which variables are
\begin{equation}
\label{root18}
A_{i,i+1}=\acute{\acute{\omega_{11}}}+\acute{\omega_{21}}x_{i+1}+\acute{\omega_{31}}x_i+\acute{\omega_{41}}x_ix_{i+1}
\nonumber
\end{equation}
\begin{equation}
\label{root19}
B_{i,i+1}=\frac{\acute{\omega_{12}}+\acute{\acute{\omega_{22}}}x_{i+1}+\acute{\omega_{32}}x_i+\acute{\omega_{42}}x_ix_{i+1}}{x_{i+1}}
\nonumber
\end{equation}
\begin{equation}
\label{root20}
C_{i,i+1}=\frac{\acute{\omega_{13}}+\acute{\omega_{23}}x_{i+1}+\acute{\acute{\omega_{33}}}x_i+\acute{\omega_{43}}x_ix_{i+1}}{x_i}
\nonumber
\end{equation}
\begin{equation}
\label{root21}
D_{i,i+1}=\frac{\acute{\omega_{14}}+\acute{\omega_{24}}x_{i+1}+\acute{\omega_{34}}x_i+\acute{\acute{\omega_{44}}}x_ix_{i+1}}{x_ix_{i+1}}
\nonumber
\end{equation}
\begin{equation}
\label{root22}
L_1=a+cx_1
\nonumber
\end{equation}
\begin{equation}
\label{root23}
L_2=\frac{b+dx_1}{x_1}
\nonumber
\end{equation}
\begin{equation}
\label{root24}
R_1=f+hx_L
\nonumber
\end{equation}
\begin{equation}
\label{root25}
R_2=\frac{g+kx_L}{x_L}
\end{equation}
Also, the largest eigenvalue of $W(s)$ is required to be
\begin{equation}
\label{root26}
\Lambda^*(s)=R_1+L_1+\Sigma_{i=1}^{L-1}A_{i,i+1}
\end{equation}

Our calculations hold for any arbitrary time-integrated observable which can be local or global. Equations \eqref{root14}-\eqref{root26} reduce to equations (9) and (10) of [\onlinecite{torkaman}] for all $x_s=1$ and $A_1$ being a time-integrated current. Therefore, [\onlinecite{torkaman}] can be considered as a special case of our calculations.

$W^{eff}(s)$ has a structure similar to \eqref{root13}. From \eqref{root113}, considering activity as a time-integrated observable, the explicit form of effective generator related to the interactions between sites $i$ and $i+1$, $\emph{h}^{eff}_{i,i+1}(s)$, is obtained
\[
\emph{h}^{eff}_{i,i+1}(s)=\left(
\begin{array}{cccc}
  \omega_{11} & \omega_{12} e^s \frac{1}{x_{i+1}} & \omega_{13} e^s \frac{1}{x_i} & \omega_{14} e^s \frac{1}{x_i x_{i+1}} \\
  \omega_{21} e^s x_{i+1} & \omega_{22} & \omega_{23}e^s \frac{x_{i+1}}{x_i} & \omega_{24} \frac{1}{x_i} e^s \\
  \omega_{31} x_i e^s & \omega_{32} e^s \frac{x_i}{x_{i+1}} & \omega_{33} & \omega_{34} e^s \frac{1}{x_{i+1}} \\
  \omega_{41} e^s x_i x_{i+1} & \omega_{42} e^s x_i & \omega_{43} e^s x_{i+1} & \omega_{44}
\end{array}\right)
\] \\
Hence, the generator of effective dynamics contains local but site-dependent interactions. It is noted that according to \eqref{root113}, for two-site creation or annihilation reaction, if $x_i=x_{i-2}$, the effective interactions are local but site-independent.
\subsection{Example:Asymmetric Zero Temperature Glauber Model}
In this section, asymmetric zero temperature Glauber model is presented as an example to show that how the general constrains lead to local but site-dependent effective interactions in the original system of hardcore particles with atypical values of activity as the time-integrated observable. In the bulk of the lattice, the dynamic of the process is $\omega_{13}=\omega_1$ and $\omega_{43}=\omega_2$, and other transition rates in the bulk of the lattice are zero. Also, particles enter and exit the left boundary with rates $\alpha$ and $\gamma$, respectively. In the right boundary, particles enter with rate $\delta$ and exit with rate $\beta$.\\
considering $x_1=1$ and constraints \eqref{root14} lead us to
 \begin{equation}
 \label{root29}
\alpha=\gamma
\nonumber
\end{equation}
\begin{equation}
\label{root30}
\delta=\beta=0
\end{equation}
and
\begin{equation}
\label{root27}
s=\ln{\frac{\omega_1+\omega_2}{\omega_1+\omega_2 x_2}}
\end{equation}
$\emph{s}$ is a set of values under which the effective stochastic generator is local but site-dependent. Also, for $2\leq i \leq L$, we have
\begin{equation}
\label{root28}
x_i=\frac{-\omega_1+x_{i-1}( \omega_1+x_2 \omega_2)}{x_{i-1} \omega_2}
\end{equation}
Solving the above equation, $x_i$ is obtained:
\begin{equation}
\label{root33}
x_i=\frac{u v (- v+(\frac{v}{u})^i(u-1)+1)}{u(\frac{v}{u})^i (u-1)+v(- v+1)}
\end{equation}
where \emph{u} and \emph{v} are
 \begin{equation}
 \label{root34}
u=\frac{\frac{\omega_1+x_2 \omega_2}{\omega_2}+\sqrt{{(\frac{\omega_1+x_2\omega_2}{\omega_2})^2}-4 \frac{\omega_1}{\omega_2}}}{2}\\
\nonumber
\end{equation}
\begin{equation}
\label{root35}
v=\frac{\frac{\omega_1+x_2 \omega_2}{\omega_2}-\sqrt{{(\frac{\omega_1+x_2\omega_2}{\omega_2})^2}-4 \frac{\omega_1}{\omega_2}}}{2}
\end{equation}
Using Vieta formula, the normalization factor of $ \langle \tilde{\Lambda}^* (s)|$, $Z$ is obtained:
\begin{equation}
\label{root36}
\prod_{i=1}^L (1+x_i)=\sum_{k=0}^L  e_k(x_1,x_2,...,x_L)
\end{equation}
$e_k$ is defined as the following:
\begin{equation}
\label{root37}
e_k(x_1,....,x_L)=\sum_{1\leq j_1< j_2< ... < j_k \leq L} x_{j_1} x_{j_2} ...x_{j_k}
\end{equation}

Since $\langle \tilde{\Lambda}^*(s)|C\rangle$ is the probability of observing the configuration $C$ at the initial time in the biased dynamics [\onlinecite{jack}], we have
\begin{equation}
\label{root38}
- \omega_1+ x_{i-1}( \omega_1+x_2 \omega_2)>0
\end{equation}
this is a necessary constraint under which $\langle \tilde{\Lambda} ^*(s) |$ is in form of  \eqref{root12}. For any positive $\omega_1$, $\omega_2$, from \eqref{root27} and \eqref{root38}, it is concluded that the effective generator is local but site-dependent for negative and small positive $s$.

Numerically, it is obtained that the unnormalized effective stationary state $|P_{TTI}\rangle$ for negative and small positive s and any arbitrary positive values of $\omega_1$ and $\omega_2$ is
\begin{equation}
\label{root207}
|P_{TTI}\rangle=\begin{pmatrix}
0\\
\vdots\\
0\\
1\\
0\\
\vdots\\
0\\
1
\end{pmatrix}
\end{equation}
The elements $2^{L-1}$th and $2^L$th of the vector are one.  The element $2^{L-1}$th is related to the configuration $\{0,1,1,....,1\}$ in which only the first site of the lattice is empty and the others are occupied. The element $2^L$th is related to the configuration $\{1,1,1,...,1\}$ in which all sites are occupied. Other elements of the vector are zero.\\
$P_{TTI}(C)$ is the probability of observing the configuration $C$ in the stationary state of the effective dynamics [\onlinecite{jack}].\\
In addition, Using
\begin{equation}
\label{root39}
P_{TTI}(C)=\langle C| \Lambda^*(s)\rangle \langle \tilde{\Lambda}^*(s)|C\rangle
\end{equation}
$\langle C| \Lambda^*(s)\rangle$ can be simply obtained [\onlinecite{jack}].\\
The largest eigenvalue is given by
\begin{equation}
\label{root31}
\Lambda^*(s)=-\alpha+\alpha e^s
\end{equation}

Through Legendre Fenchel transformation, the rate function is [\onlinecite{touchette}].
\begin{align}
\label{root32}
I(J)=J \ln{\frac{J}{\alpha}}+\alpha-J
\end{align}

Finally, from \eqref{root113}, the explicit form of $h^{eff}_{i,i+1}$ is obtained. For example, $ h^{eff}_{12}$ and $h^{eff}_{23}$ which are the effective generators of sites one, two and two, three, respectively, are obtained:
\[
h^{eff}_{1,2}=\left(
  \begin{array}{cccc}
    0 & 0 & \omega_1 \frac{\omega_1+\omega_2}{\omega_1+\omega_2 x_2} & 0 \\
    0 &0 & 0 & 0 \\
    0 & 0 & - \omega_1- \omega_2 & 0 \\
    0 & 0 &  \omega_2 \frac{\omega_1+\omega_2}{\omega_1+\omega_2 x_2} x_2 & 0 \\
  \end{array}\\
\right)
\]
\[
h^{eff}_{2,3}=\left(
  \begin{array}{cccc}
    0 & 0 & \frac{\omega_1}{x_2}  \frac{\omega_1+\omega_2}{\omega_1+\omega_2 x_2} & 0 \\
    0 &0 & 0 & 0 \\
    0 & 0 & - \omega_1- \omega_2 & 0 \\
    0 & 0 &   \frac{\omega_1+\omega_2}{x_2(\omega_1+\omega_2 x_2)}(-\omega_1+x_2 \omega_1+x_2^2 \omega_2)  & 0 \\
  \end{array}\\
\right)
\]
It can be seen that the effective transition rates under certain constraints are local but site-dependent.

\section{effective generator with site-dependent interactions in a noninteracting particle system}
We consider a noninteracting particle system with nearest-neighbor interactions on an \emph{L}-site one-dimensional lattice with open boundaries, from there particles can enter or exit the lattice, The dynamical rules in the original process is
\begin{center}
$(n_i)(n_{i+1})\rightarrow(n_i-1)(n_{i+1}+1)$~~~ with rate $\omega_1$ \\
$(n_i)(n_{i+1})\rightarrow(n_i+1)(n_{i+1}-1)$~~~ with rate $\omega_2$\\
$(n_i)(n_{i+1})\rightarrow(n_i+1)(n_{i+1})$~~~~~~~ with rate $\omega_3$\\
$(n_i)(n_{i+1})\rightarrow(n_i-1)(n_{i+1})$~~~~~~~ with rate $\omega_4$\\
$(n_i)(n_{i+1})\rightarrow(n_i)(n_{i+1}+1)$~~~~~~~ with rate $\omega_5$\\
$(n_i)(n_{i+1})\rightarrow(n_i)(n_{i+1}-1)$~~~~~~~ with rate $\omega_6$\\
$(n_i)(n_{i+1})\rightarrow(n_i-1)(n_{i+1}-1)$~~~ with rate $\omega_7$\\
$(n_i)(n_{i+1})\rightarrow(n_i+1)(n_{i+1}+1)$~~~ with rate $\omega_8$\\
\end{center}
in which $n_i$ denotes the number of particles on site $i$.\\
In the boundaries, particles can enter and exit the left boundary with rates $\alpha$ and $\gamma$, respectively, and in the right boundary with rates $\delta$ and $\beta$, respectively.\\
Now, we investigate constraints under which the effective dynamics of a noninteracting particle system with an atypical value of an arbitrary time-integrated observable contains local (short-range) but site-dependent interactions. Like the case of hardcore particle system, the left eigenvector corresponding to the largest eigenvalue of $W(s)$ is considered in a product form.
\begin{equation}
\label{root40}
\langle\tilde{\Lambda}^*(s)|=\langle{y_1}|\otimes\langle{y_2}|\otimes...\otimes\langle{y_L}|
\end{equation}
Considering an arbitrary time-integrated observable whether depends on the transition between consecutive configurations or configurations that
the system visits during the observation time, $W(s)$ is written through Doi-Peliti formalism [\onlinecite{garrahan}]:
\begin{equation}
\begin{split}
\label{root42}
\emph{W(s)}&=\Sigma_{i=1}^{\emph{L}-1}\acute{\omega_1} a_{i} a_{i+1}^\dagger+\acute{ \omega_2} a_{i}^\dagger a_{i+1}+\acute{\omega_7} a_i a_{i+1}+\acute{\omega_8} a_{i}^\dagger a_{i+1}^\dagger\\&~~~- \acute{\acute{\omega_1}} n_i-\acute{\acute{\omega_2}} n_{i+1}-\acute{\acute{ \omega_7}} n_i n_{i+1}- \acute{\acute{\omega_8}}+\Sigma_{i=1}^{\emph{L}}\acute{\omega_3} a_i^\dagger+\\&~~~ \acute{\omega_4} a_i-\acute{\acute{ \omega_3}}-\acute{\acute{\omega_4}}  n_i+\Sigma_{i=1}^{\emph{L}-1}\acute{\omega_5} a_{i+1}^\dagger+\acute{\omega_6} a_{i+1}-\acute{\acute{\omega_5}}-\\&~~~ \acute{\acute{\omega_6}} n_{i+1}+\acute{\alpha} a_1^\dagger+ \acute{\gamma} a_1-\acute{\acute{\alpha}}- \acute{\acute{\gamma}} n_1+ \acute{\beta} a_L+\acute{\delta} a_L^\dagger- \acute{\acute{\beta}} n_L\\&~~~- \acute{\acute{\delta}}
\end{split}
\end{equation}
where $a^\dagger$ and $a$ are infinite dimensional particle creation and annihilation matrices, respectively. $\acute{\omega_i}$s and $\acute{\acute{\omega_i}}$s are non-diagonal and diagonal elements of $W(s)$, respectively. They are obtained through \eqref{root106} or \eqref{root107} based on the type of the time-integrated observable that we consider in the calculations. \\
Considering \eqref{root111}, certain constraints are obtained under which $\langle \tilde{\Lambda}^*(s)|$ is the left eigenvector corresponding to the largest eigenvalue of $W(s)$
\begin{equation}
\label{root47}
 \acute{\omega_1} y_{i+1}-(\acute{\acute{\omega_1}}+\acute{\acute{\omega_4}}+\acute{\acute{\omega_2}}+\acute{\acute{\omega_6}})y_i +\acute{\omega_2}y_{i-1}= - \acute{\omega_4}-\acute{\omega_6}
 \tag{33-a}
 \end{equation}
 \begin{equation}
 \label{root48}
 \acute{\omega_7} y_i^{-1} y_{i+1}^{-1}=\acute{\acute{\omega_7}}
 \tag{33-b}
 \end{equation}
 \begin{equation}
 \label{root49}
 \acute{\omega_1}y_2-(\acute{\acute{\omega_1}}+\acute{\acute{\omega_4}}+\acute{\acute{\gamma}})y_1=- \acute{\omega_4}- \acute{\gamma}
 \tag{33-c}
 \end{equation}
 \begin{equation}
 \label{root50}
 \acute{\omega_2}y_{L-1}- (\acute{\acute{\omega_2}}+\acute{\acute{\omega_6}}+\acute{\acute{\beta}}+\acute{\acute{\omega_4}})y_L=- \acute{\omega_6}-\acute{\omega_4}-\acute{\beta}
 \tag{33-d}
\end{equation}
From \eqref{root48}, it is clear that for a stochastic process which contains a two-site creation or annihilation reaction, the effective dynamics of the system can never be site-dependent. So, since we investigate constraints under which effective dynamics contains site-dependent interactions, we ignore these two reactions and continue our calculations with the remaining ones. Certain constraints are obtained that under which $\langle \tilde{\Lambda}^*(s)|$ is the left eigenvector corresponding to the largest eigenvalue of the generator of biased dynamics:
\begin{equation}
\label{root58}
b_+ y_{i+1}+b_i y_i+ b_- y_{i-1}=b_0
\tag{34-a}
\end{equation}
\begin{equation}
\label{root59}
f_2 y_2+f_1y_1=f_0
\tag{34-b}
\end{equation}
\begin{equation}
\label{root60}
g_L y_L+g_{L-1}y_{L-1}=g_0
\tag{34-c}
\end{equation}
In \eqref{root58}, $b_+=\acute{\omega_1}$, $b_i=-(\acute{\acute{\omega_1}}+\acute{\acute{\omega_2}}+\acute{\acute{\omega_4}}+\acute{\acute{\omega_6}})$, $b_-=\acute{\omega_2}$ and $b_0=-(\acute{\omega_4}+\acute{\omega_6} )$. In \eqref{root59}, $f_2=\acute{\omega_1}$, $f_1=-(\acute{\acute{\omega_1}}+\acute{\acute{\omega_4}}+\acute{\acute{\gamma}})$ and $f_0=-(\acute{\omega_4}+\acute{\gamma})$. In \eqref{root60}, $ g_L=-(\acute{\acute{\omega_2}}+\acute{\acute{\omega_4}}+\acute{\acute{\omega_6}}+\acute{\acute{\beta}})$, $g_{L-1}=\acute{\omega_2}$ and $g_0=-(\acute{\omega_4}+\acute{\omega_6}+\acute{\beta})$.\\
Using the ansatz
\begin{equation}
\label{root3335}
y_i=A Z_1^i+B Z_2^i+Z_0
\tag{35}
\end{equation}
after a slight simplification, from \eqref{root58}, we have:
\begin{equation}
\label{root62}
Z_{1,2}=\frac{- b_i\pm\sqrt{b_i^2- 4 b_+ b_-}}{2 b_+}
\tag{36-a}
\end{equation}
\begin{equation}
\label{root63}
Z_0=\frac{b_0}{ b_++b_0+b_-}
\tag{36-b}
\end{equation}
From \eqref{root59} and \eqref{root60}, A and B are determined:
\begin{widetext}
\begin{equation}
\label{root64}
A=-\frac{(-g_0+(g_{-1+L}+g_L)Z_0)(f_1+f_2 Z_2) Z_2+(f_0-(f_1+f_2)Z_0)Z_2 ^ {-1+L}(g_{-1+L}+g_L Z_2)}{Z_1 ^{-1+L}(g_{-1+L}+g_L Z_1) Z_2 (f_1 +f_2 Z_2)- Z_1 (f_1 + f_2 Z_1)Z_2 ^{-1+L} (g_{-1+L}+g_L Z_2)}
\tag{37-a}
\end{equation}
\begin{equation}
\label{root65}
B=\frac{(-(g_0- (g_{-1+L}+g_L)Z_0)Z_1 ^2 (f_1+f_2 Z_1)+(f_0 - (f_1 +f_2) Z_0)Z_1 ^L (g_{-1+L}+g_LZ_1))Z_2}{Z_1 ^L (g_{- 1+L}+g_L Z_1)Z_2 ^2 (f_1+f_2 Z_2)- Z_1^2 (f_1 +f_2 Z_1)Z_2^ L (g_{-L+1}+g_L Z_2)}
\tag{37-b}
\end{equation}
\end{widetext}
From \eqref{root62}, the valid area for calculations is obtained:
\begin{equation}
\label{root70}
(\acute{\acute{\omega_1}}+\acute{\acute{\omega_2}}+\acute{\acute{\omega_4}}+\acute{\acute{\omega_6}})^2\geq 4 \acute{\omega_1} \acute{\omega_2}
\tag{38}
\end{equation}
The largest eigenvalue of $W(s)$ is required to be
\begin{equation}
\label{root66}
\Lambda^*(s)=\sum_{i=1}^L \acute{\omega_3}y_i-\acute{\acute{\omega_3}}+\sum_{i=1}^{L-1} \acute{\omega_5}y_{i+1}-\acute{\acute{\omega_5}}+\acute{\alpha} y_1-\acute{\acute{\alpha}}+\acute{\delta}y_L-\acute{\acute{\delta}}
\tag{39}
\end{equation}
Using \eqref{root112}, the explicit form of effective stochastic generator is obtained:
\begin{equation}
\begin{split}
\label{root69}
W^{eff}(s)&=\Sigma_{i=2}^{\emph{L}-1}\acute{\omega_1} y_{i+1} y_i^{- 1} a_i a_{i+1}^\dagger-\acute{\acute{\omega_1}}n_i+\acute{\omega_2} a_i a_{i-1}^\dagger y_{i-1}y_i^{- 1}\\&-\acute{\acute{\omega_2}} n_i +\acute{\omega_1} y_{2} y_1^{- 1} a_1 a_{2}^\dagger-\acute{\acute{\omega_1}}n_1+\acute{\omega_2} a_L a_{L-1}^\dagger y_{L-1}y_L^{- 1}\\&-\acute{\acute{\omega_2}} n_L+ \acute{\omega_3} a_i^\dagger y_i- \acute{\acute{\omega_3}}+ \acute{\omega_4} a_i y_i^{- 1}- \acute{\acute{\omega_4}} n_i+\acute{\omega_3} a_1^\dagger y_1\\&- \acute{\acute{\omega_3}}+\acute{\omega_3} a_L^\dagger y_L- \acute{\acute{\omega_3}}+ \acute{\omega_4} a_1 y_1^{- 1}- \acute{\acute{\omega_4}} n_1+ \acute{\omega_4} a_L y_L^{- 1}\\&- \acute{\acute{\omega_4}} n_L+ \acute{\omega_5} y_i a_i^\dagger-\acute{\acute{\omega_5}}+ \acute{\omega_5} y_L a_L^\dagger-\acute{\acute{\omega_5}}+\acute{\omega_6}a_i y_i^{- 1}\\& - \acute{\acute{\omega_6}} n_i+\acute{\omega_6}a_L y_L^{- 1} - \acute{\acute{\omega_6}} n_L+\acute{\alpha}y_1 a_1^\dagger-\acute{\acute{\alpha}}+\acute{\gamma}a_1 y_1^{- 1}\\&- \acute{\acute{\gamma}}n_1+\acute{\beta}y_L^{- 1} a_L- \acute{\acute{\beta}}n_L+\acute{\delta} y_L a_L^\dagger-\acute{\acute{\delta}}-(\sum_{i=1}^L \acute{\omega_3}y_i\\&-\acute{\acute{\omega_3}}+\sum_{i=1}^{L-1} \acute{\omega_5}y_{i+1}-\acute{\acute{\omega_5}}+\acute{\alpha} y_1-\acute{\acute{\alpha}}+\acute{\delta}y_L-\acute{\acute{\delta}})
\end{split}
\tag{40}
\end{equation}
Hence, the effective generator under constraints is local (short-range) and site-dependent.

It is worth mentioning that if the constraints obtained are applied on zero-range process with open boundary and local current between left reservoir and the first site, as studied in paper [\onlinecite{hirschberg}], the above constraints reduce to those in [\onlinecite{hirschberg}]. it means that [\onlinecite{hirschberg}] is a special case of our calculations.
\subsection{Example:A-model with diffusion}
As an illustration,  A-model [\onlinecite{garrahan}] with diffusion on an \emph{L}-site one-dimensional lattice with open boundaries is presented to show that how constrains which is obtained in previous section result in local but site-dependent effective interactions in the original system with an atypical value of activity as the time-integrated observable. The original dynamics is the same as the previous section with $\omega_7=0$ and $\omega_8=0$.\\
 From \eqref{root70}, the domain of $s$ is obtained:
\begin{equation}
\label{root74}
- \ln\frac{\omega_1+\omega_2+\omega_4+\omega_6}{2 \sqrt{\omega_1\omega_2}} \leq s\leq\ln\frac{\omega_1+\omega_2+\omega_4+\omega_6}{2 \sqrt{\omega_1\omega_2}}
\tag{41}
\end{equation}
and $\langle \tilde{\Lambda}^*(s)|$ in \eqref{root40} is obtained from \eqref{root3335} to \eqref{root65}.
Also, the largest eigenvalue of $W(s)$ is
\begin{equation}
\begin{split}
\label{root75}
\Lambda^*(s)&=\sum_{i=1}^L \omega_3 (e^s y_i-1)+\sum_{i=1}^{L-1} \omega_5 (e^s y_i-1)+\alpha (e^s y_1-1)\\&+\delta (y_L e^s-1)
\end{split}
\tag{42}
\end{equation}
From \eqref{root112}, the explicit form of effective generator of A-model with diffusion is obtained:
\begin{equation}
\begin{split}
\label{root69}
W^{eff}(s)&=\Sigma_{i=2}^{\emph{L}-1}\omega_1 e^s y_{i+1} y_i^{- 1} a_i a_{i+1}^\dagger-\omega_1n_i+\omega_2 e^s a_i a_{i-1}^\dagger\\&~~~ y_{i-1}y_i^{- 1}-\omega_2 n_i +\omega_1 e^s y_{2} y_1^{- 1} a_1 a_{2}^\dagger-\omega_1n_1+\omega_2 e^s\\&~~~ a_L a_{L-1}^\dagger y_{L-1}y_L^{- 1}-\omega_2 n_L+ \omega_3 e^s a_i^\dagger y_i- \omega_3+ \omega_4 e^s\\&~~~ a_i y_i^{- 1}- \omega_4 n_i+\omega_3 e^s a_1^\dagger y_1- \omega_3+\omega_3 e^s a_L^\dagger y_L- \omega_3\\&~~~+ \omega_4 e^s a_1 y_1^{- 1}- \omega_4 n_1+\omega_4 e^s a_L y_L^{- 1}- \omega_4 n_L+ \omega_5\\&~~~ e^s y_i a_i^\dagger-\omega_5+ \omega_5 e^s y_L a_L^\dagger-\omega_5+\omega_6 e^s a_i y_i^{- 1} - \omega_6 n_i\\&~~~+\omega_6 e^s a_L y_L^{- 1} - \omega_6 n_L+\alpha e^s y_1 a_1^\dagger-\alpha+\gamma e^s a_1 y_1^{- 1}\\&~~~- \gamma n_1+\beta e^s y_L^{- 1} a_L- \beta n_L+\delta e^s y_L a_L^\dagger-\delta-(\sum_{i=1}^L \\&~~~\omega_3 e^s y_i-\omega_3+\sum_{i=1}^{L-1} \omega_5 e^s y_{i+1}-\omega_5+\alpha e^s y_1-\alpha+\\&~~~\delta e^s y_L-\delta)
\end{split}
\tag{43}
\end{equation}
So, the effective generator of A-model with diffusion under the constraints is local and site-dependent
\section{Conclusion}
It is known that the particle interactions in the stochastic effective generator of a particle system which is conditioned on an atypical value of a time-integrated observable is usually non-local. In this paper we have shown that there are certain cases for which these effective interactions are local (short-range). The systems that we have considered consist of classical particles hoping on a one-dimensional lattice of finite size with open boundaries. Two types of particles have also been considered: the hardcore particles and noninteracting particles. The hardcore particles have hard-core interactions so that each lattice site can be occupied by at most one of them while noninteracting particles can accumulate on a single lattice site.

It turns out that the left eigenvector corresponding to the largest eigenvalue of $W(s)$ plays an important role in the structure of the effective generator; therefore, by conjecturing different types for this vector, in comparison to the most simplest case introduced in [\onlinecite{torkaman}], we have been able to find certain constraints, which define certain process, for which the stochastic effective generator can be calculated exactly. The key point to this achievement was, in fact, considering a product form for the left eigenvector corresponding to the largest eigenvalue of $W(s)$.

It is worth mentioning that our calculations are true for any arbitrary time-integrated observable, such as particle current, which can be local or global. On the other hand, the physical observable can be of the type which depends on the transition between consecutive configurations or configurations that system visits during the observation time.
\bibliography{aipsamp}

\end{document}